\documentclass[usenatbib]{mn2e}
\usepackage{epsfig}
\usepackage{subfigure}
\usepackage{amssymb}
\usepackage{color}
\usepackage{natbib}
\usepackage{xcolor}

\newcommand{\apj}{ApJ}

\newcommand{\apjl}{ApJL}
\newcommand{\apjs}{ApJS}
\newcommand{\mnras}{MNRAS}

\newcommand{\nat}{Nature}

\newcommand{\pasp}{PASP}

\newcommand{\bn}{\begin{enumerate}}
\newcommand{\en}{\end{enumerate}}
\newcommand{\bi}{\begin{itemize}}
\newcommand{\ei}{\end{itemize}}

\newcommand{\Msun}{M_{\odot}}
\newcommand{\Mstar}{M_{\star}}

\newcommand{\rhostar}{\rho_{\rm *}}
\newcommand{\himsun}{h^{-1}\Msun}

\title{On the inconsistency between the estimates of cosmic star formation rate and stellar mass density of high redshift galaxie}
\author[Choi \& Nagamine]
{Jun-Hwan Choi\thanks{Current address:  Department of Physics \& Astronomy, University of Kentucky, Lexington, KY 40506-0055, U.S.A.} \thanks{Email: jhchoi@pa.uky.edu}, Kentaro Nagamine
\vspace{0.2cm}\\
Department of Physics and Astronomy, University of Nevada, Las Vegas,
4505 S. Maryland Pkwy, Las Vegas, NV, 89154-4002, U.S.A.}

\begin{document}

\maketitle

\begin{abstract}
There are mainly two different approaches to measure the cosmic star formation history: direct star formation rate density (SFRD) and stellar mass density $\rhostar$ as functions of redshift. 
Compilations of current observations seem to show a disparity in the two quantities, in the sense that the integral of SFRD is higher than the observed $\rhostar$ (after considering gas recycling). 
Using cosmological smoothed particle hydrodynamics simulations based on the concordance $\Lambda$ cold dark matter model, we show that the two quantities become more consistent with each other when we consider the observed galaxy mass limit.
The comparison between simulations and (dust corrected) observed cosmic SFRD shows a good agreement, while the observed $\rhostar$ is significantly lower than the simulation results.  
This can be reconciled if the current high-$z$ galaxy surveys are missing faint low-mass galaxies due to their flux limit. 
Our simulated GSMFs have steep low-mass end slopes of $\alpha \lesssim -2$ at $z>3$, and when these numerous low-mass galaxies are included, the total $\rhostar$ matches with the integral of SFRD.
\end{abstract}


%% ==================================================================

\section{Introduction}
\label{sec:intro}

The cosmic star formation (SF) history is a fundamental quantity that illustrates the buildup of galaxies.
It can be estimated by direct measurement of SFRD as a function of redshift through rest-frame H$\alpha$, UV, and far-IR emission with dust extinction corrections \citep[e.g.,][]{Hopkins.Beacom:06}.
The stellar mass density ($\rhostar$) at various redshifts is an indirect measurement of the cosmic SF history, because it is indeed the integral of SFRD, after considering the recycling of gas into interstellar medium.
Given this simple relationship, in principle the measurements of the two quantities should be consistent with each other.  
However, in this paper we show that this is not the case for current observational estimates, and other authors have reached similar conclusions \citep{Nagamine.etal:04, Ouchi.etal:04, vanDokkum:08, Wilkins.etal:08, Dave:08}. 

Over the past decade, large samples of color selected high-$z$ galaxies such as Lyman break galaxies (LBGs) have enabled us to measure the cosmic SFRD at $3 \lesssim z \lesssim 6$ from rest-frame UV luminosity density \citep[e.g.,][]{Steidal.etal:99}.
A number of near-infrared observations have also constrained the galaxy stellar mass function (GSMF) at $z\lesssim 4$ \citep[e.g.,][]{Marchesini.etal:09}.
Furthermore, the advent of the {\it Wide Field Camera 3} ({\it WFC3}) on board the {\it Hubble Space Telescope} ({\it HST}) has dramatically improved our ability to measure the rest-frame UV light from galaxies at $z \gtrsim 6$.
Combined with the measurement of rest-frame optical light by the {\it Infrared Array Camera} ({\it IRAC}) on {\it Spitzer Space Telescope}, it is now possible to measure the characteristic stellar mass of galaxies at $z \gtrsim 6$ using stacked spectra \citep{Yan.etal:09,Bouwens.etal:10,Labbe.etal:10,Gonzalez.etal:10}.

%Table loc
\begin{table*}
\begin{center}
\begin{tabular}{ccccccc}
\hline
Name &  Box-size & ${N_{\rm p}}$ & $m_{\rm DM}$ & $m_{\rm gas}$ & $\epsilon$ & $z_{\rm end}$ \\
\hline
\hline
N216L10    & 10.0  & $2\times 216^3$  & $5.96 \times 10^6$ & $1.21 \times 10^6$ &  1.85  & 2.75 \cr
\hline
N400L34   & 33.75 & $2\times 400^3$  & $3.49 \times 10^7$ & $7.31 \times 10^6$ &  3.38  & 1.0 \cr
\hline
N600L100   & 100.0 & $2\times 600^3$  & $2.70 \times 10^8$ & $5.66 \times 10^7$ &  4.30  & 0.0 \cr
\hline
\label{table:sim}
\end{tabular}
\caption{
The three different resolution and volume simulations employed in this {\it Letter}.
The box-size is given in units of $h^{-1}$Mpc, ${N_{\rm p}}$ is the particle number of dark matter and initial gas (hence $\times\, 2$), $m_{\rm DM}$ and $m_{\rm gas}$ are the masses of dark matter and initial gas particles in units of $\himsun$, respectively, $\epsilon$ is the comoving gravitational softening length in units of $h^{-1}$kpc, and $z_{\rm end}$ is the ending redshift of the simulation.
Note that the mass of star particle is a half of the initial gas particle.
The value of $\epsilon$ is a measure of spatial resolution.
The name of each simulation is based on the particle count and its box size. 
}
\end{center}
\end{table*}

However, the sources at $z>6$ are very faint, and so far we have only detected the massive end of GSMF.
In addition, there are still significant uncertainties in the estimates of SFRD and $\rhostar$. 
Given this situation, it would be useful to obtain predictions on SFRD and $\rhostar$ from theoretical models.  
In particular, cosmological hydrodynamic simulations have been widely used to investigate cosmic star formation \citep[e.g.,][]{Cen.Ostriker:92,Katz.etal:96,Springel.Hernquist:03_SFR,Nagamine.etal:06,Dave:08,Schaye.etal:10}.

In this {\it Letter}, we focus on the cosmic SF history at $z>2$ using cosmological simulations.
The remaining of this paper is organized as follows.
In Section~\ref{sec:method}, we describe the cosmological hydrodynamic simulations.
In Section~\ref{sec:MF}, we construct the composite GSMF by combining the samples of galaxies from simulations with different resolution and volumes.
We study the cosmic SFR and $\rhostar$ evolution in Section~\ref{sec:SF}.
We summarise and discuss our findings in Section~\ref{sec:summary}.

\section{Numerical Simulations}
\label{sec:method}

We use the modified version of the tree-particle-mesh smoothed particle hydrodynamics (SPH) code GADGET-3 \citep[originally described in][]{Springel:05}.
Our conventional code includes radiative cooling by H, He, and metals \citep{Choi.Nagamine:09}, heating by a uniform UV background of a modified \citet{Haardt.Madau:96} spectrum \citep{Katz.etal:96,Dave.etal:99}, star formation, its feedback, a phenomenological model for galactic winds, and a sub-resolution model of multiphase ISM \citep{Springel.Hernquist:03}.
For the star formation model, we use the ``Pressure model'' which reduces the high-$z$ SFRD \citep{Schaye.DallaVecchia:08,Choi.Nagamine:10} relative to the previous model by \citet{Springel.Hernquist:03}.
For the galactic outflow model, we use the Multicomponent Variable Velocity wind model developed by \citet{Choi.Nagamine:11}.
We adopt the following cosmological parameters that are consistent with the WMAP best-fit values \citep{Komatsu.etal:11}: $\Omega_m = 0.26$, $\Omega_{\Lambda} = 0.74$, $\Omega_b = 0.044$, $h=0.72$, $n_{s}=0.96$, and $\sigma_{8}=0.80$.
To identify galaxies in simulations, we use simplified variant of the SUBFIND algorithm \citep{Springel.etal:01,Choi.Nagamine:09}.

Although cosmological simulations have been widely used to study galaxy formation, there are inevitable limitations due to resolution and box size. 
One critical limitation is that the low-mass galaxies are not captured in a low-resolution simulation with a large box size, and the high mass galaxies are missed in a small box size simulation due to limited box size. 
To alleviate this problem, we construct composite GSMFs in a wide range of galaxy stellar mass using a few simulations with different resolution and volumes (see Table~\ref{table:sim}).

An important point to note is that the SFRD is one of the most direct output of our ab initio cosmological simulations; our simulation follows the gas and dark matter dynamics within the framework of $\Lambda$CDM model, and converts part of the gas into star particles when the SF threshold density is satisfied at every time step. Therefore the SFRD is the most direct outcome of gas dynamics, and no uncertain conversion is necessary to obtain SFRD from our simulations. 
This is why our simulations are helpful to resolve this problem between SFRD and $\rhostar$.

\section{Galaxy stellar mass function}
\label{sec:MF}

\begin{figure*}
\centerline{\includegraphics[width=0.9\textwidth,angle=0] {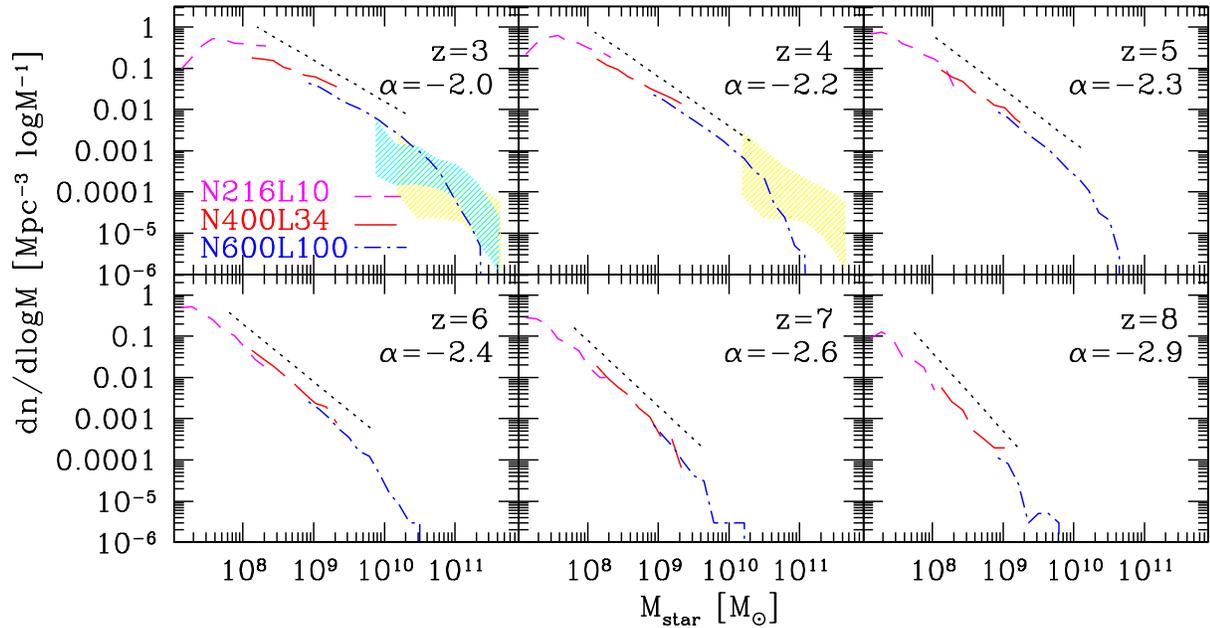}}
\caption{
Simulated GSMFs at $z$=3, 4, 5, 6, 7, and 8, from our three simulations.
The low mass range ($\Mstar < 1 \times 10^8 \himsun$) is represented by the N216L10 run, the intermediate mass range by the N400L34 run, and the high mass range ($\Mstar > 1 \times 10^9 \himsun$) by the N600L100 run. 
The shaded regions show the range of observations at $2< z <3$ (blue shade) and at $3< z <4$ (yellow shade) \citep{Marchesini.etal:09}. 
The black dotted lines show the low-mass end slopes of $\alpha$ indicated in each panel.  
The simulated GSMFs show good agreement with observational data, and they have significantly steep low-mass end slopes in the mass-scales where the observations are not available due to faintness of those sources. 
}
\label{fig:MF}
\end{figure*}

Figure~\ref{fig:MF} shows the composite GSMFs for different redshifts.
Three simulations with different resolutions and volumes (N216L10, N400L34, and N600L100 runs) are used to construct the composite GSMFs, which cover the galaxy stellar mass range of $10^7 \himsun < \Mstar < 10^{11} \himsun$. 
The lowest mass galaxies are represented by the N216L10 run, the intermediate mass galaxies by the N400L34 run, and the most massive galaxies by the N600L100 run.
The covered mass range for each run is determined by the following two criteria: 
First, the lowest $\Mstar$ of a given simulation is equivalent to that of 32 star particles, which is $\sim 10^7 \himsun$, $\sim 10^8 \himsun$, and $\sim 10^9 \himsun$ for the N216L10, N400L34, and N600L100 runs, respectively. 
Second, if the GSMFs overlap between different runs, the one from a larger volume run is used, because a smaller volume run tend to underestimate the number density of galaxies at the massive end owing to the limited box size.  
Figure~\ref{fig:MF} shows marginal overlaps between two simulations for presentation purposes.
Three segments of GSMF from three simulations smoothly form one mass function. 

The simulated GSMF shows a reasonable agreement with observations (yellow and blue shade) at $2\lesssim z \lesssim 4$, although the simulated GSMF is slightly lower than the observations at the very massive end of $\Mstar > 10^{11} \himsun$. 
The number density of these high-mass galaxies is so low that there are only about $1-10$ galaxies in a volume of $(100 h^{-1}{\rm Mpc})^3$, therefore this discrepancy could be due to the cosmic variance.
Another possible reason is that the lower mass galaxies are scattered up into the more massive bins due to observational errors, causing it to be higher than the intrinsic simulated GSMF. 

A marked feature in the simulated GSMF is the very steep slope at the low-mass end. 
When fitted with a Schechter function, the simulated composite GSMF has a faint-end slope of $\alpha \sim -2$ at $z=3$ for the mass range of $10^8 \himsun < \Mstar < 10^{10}\himsun$, while the observations suggest $\alpha \sim -1.6$ based on the data at $\Mstar \gtrsim 10^{10}\himsun$ \citep[e.g.][]{Marchesini.etal:09}.
The simulation suggests that a large fraction of stars is located in the galaxies with $\Mstar < 10^{10} \himsun$ at $z=3$.
We find that 22\% of total $\rhostar$ is located in the galaxies with $\Mstar > 10^{10} \himsun$, 34\% in $10^{9} \himsun < \Mstar < 10^{10} \himsun$, 31\% in $10^{8} \himsun < \Mstar < 10^{9} \himsun$, and 13\% in $10^{7} \himsun < \Mstar < 10^{8} \himsun$ for the simulated GSMF at $z=3$.
In general, current galaxy surveys cannot detect galaxies with $\Mstar < 10^{10} \himsun$ due to their flux limit at $z > 3$, and deeper observations are needed to probe the GSMF at lower masses. 

Another notable feature of simulated GSMF is that its low-mass slope becomes steeper as $z$ increases.
For example, the slope is $\alpha \sim -2.4$ at $z=6$, and $\alpha \sim -2.9$ at $z=8$.  
We perform more rigorous $\chi^2$ fitting with a Schechter function in a separate publication \citep{Jaacks.etal:2011}, which still shows $\alpha \lesssim -2$.
This steepness of simulated GSMF has significant consequences when one measures the global quantities such as cosmic SFRD and $\rhostar$.
We note that the integration of GSMF to $ \Mstar $ = 0 will diverge if $\alpha < -2$, and it requires a low-mass limit to compute the total $\rhostar$.  
In our simulations, the lowest galaxy mass is determined by the atomic cooling limit of $T\sim 10^4$\,K, which corresponds to $\Mstar \sim 10^7 \Msun$.  
Therefore the summation of simulated galaxy stellar mass stops at $ \Mstar \sim 10^{7} \himsun$.

\section{High-z SFRD and $\rhostar$}
\label{sec:SF}

\begin{figure*}
\centerline{
  \includegraphics[width=0.45\textwidth,angle=0] {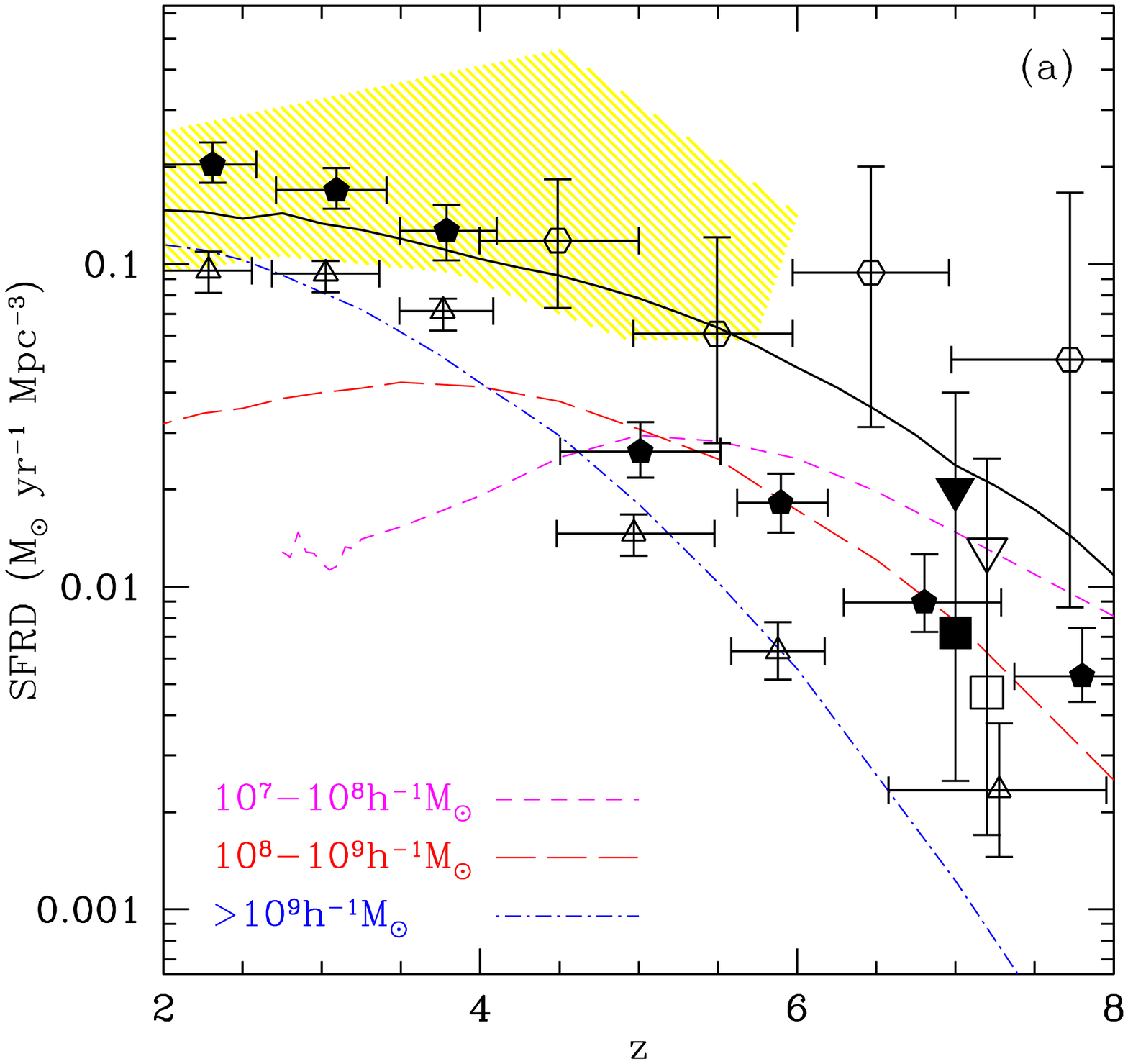}
  \includegraphics[width=0.45\textwidth,angle=0] {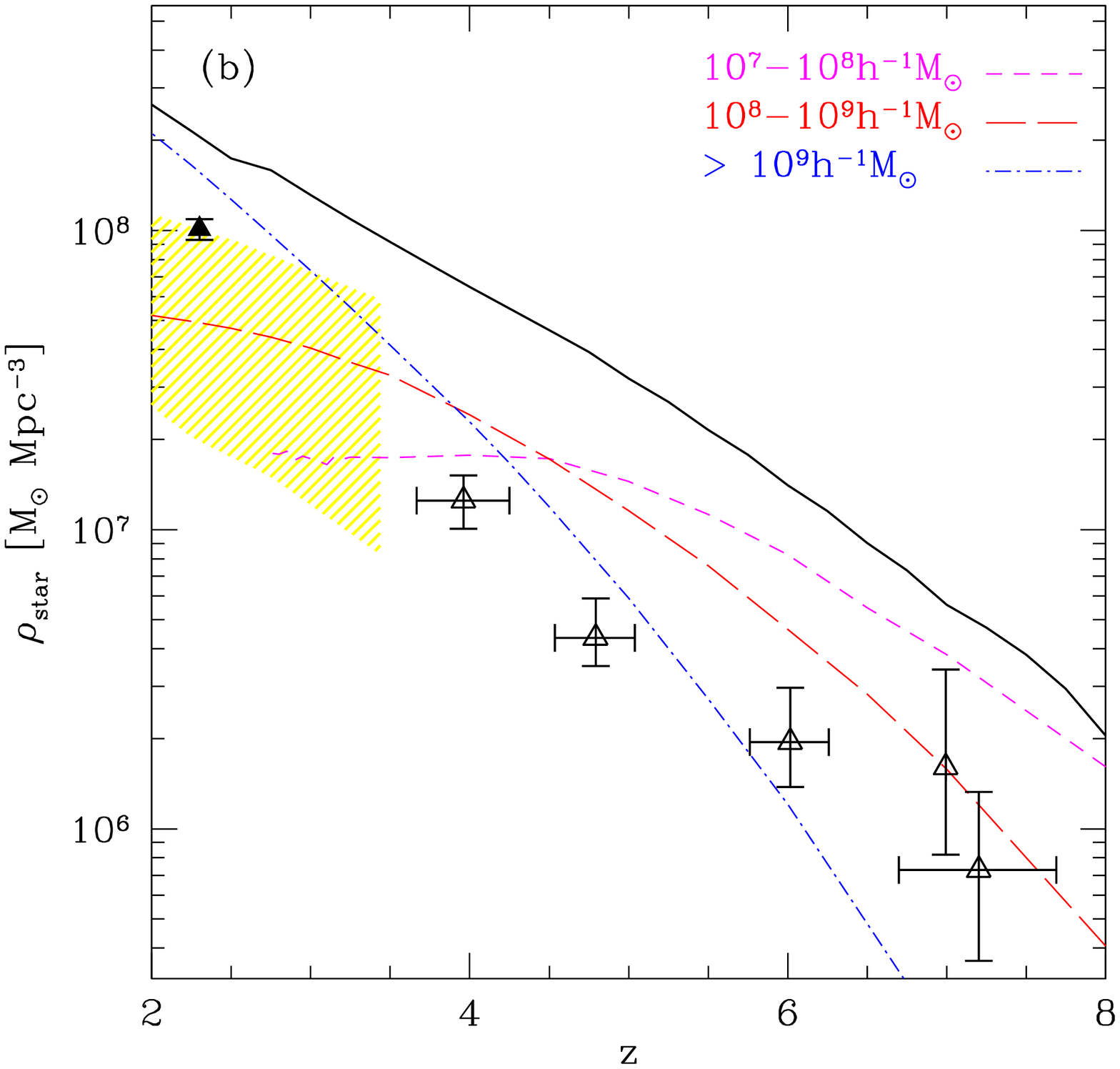}
}
\caption{
Cosmic SFRD ({\it left}) and $\rhostar$ ({\it right}) as functions of redshift.
In each panel, the contribution from three different stellar mass ranges ($10^7 - 10^8 \himsun$, $10^8 - 10^9 \himsun$, and $ > 10^9 \himsun$) is shown.
The black solid line in each panel show the {\it total} cosmic SFRD and $\rhostar$ from all galaxies in three different mass ranges.
{\it Left:} The yellow shade is the locus of the observed data compiled by \citet{Nagamine.etal:06}, corrected for the Salpeter IMF. 
The black open pentagons with error bars are the observational estimates of SFRD using the high-$z$ from $\gamma$-ray burst data from \citet{Kistler.etal:09}.
The black open triangles with error bars are the compilation of observational estimates of SFRD from \citet{Gonzalez.etal:10} for galaxies with $> 0.3 L^{\star}_{z=3}$. 
The black solid pentagons with error bars are the compilation of observational estimates of SFRD from \citet{Bouwens.etal:10b} for galaxies with $> 0.05 L^{\star}_{z=3}$ with dust correction. 
The black squares and inverted-triangles at $z\sim 7$ are the observational estimates of SFRD with (solid symbols) and without (open symbols) dust correction from \citet{Ouchi.etal:09}; the squares are the SFRD integrated down to $L \sim 0.1 L_{\star}$ and the inverted-triangles are the SFRD integrated down $L = 0$, which agrees well with our total sum (solid black line). 
The black solid line shows good agreement with the observed range of SFRD shown by the yellow and other observational data points.
{\it Right:} The yellow shade is the estimated $\rhostar$ by integrating the GSMFs over the stellar mass range of $\Mstar > 10^8\himsun$ \citep{Marchesini.etal:09}.
The GSMFs in \citet{Marchesini.etal:09} are established by fitting the \citet{Schechter:76} function to the observed galaxies with $\Mstar \gtrsim 10^{10} \himsun$.
The black open triangles with error bars are the observed data from \citet{Gonzalez.etal:10}.
The black solid triangle at $z\sim 2$ with error bars is the observed data from \citet{Reddy.Steidel:09} for $L>0.083 L^{\star}_{z=2.3}$. 
The black solid line is clearly higher than the observational estimate by \citet{Marchesini.etal:09}.
All observational data are corrected for the \citet{Salpeter:55} IMF.
}
\label{fig:sf}
\end{figure*}

Figure~\ref{fig:sf} shows the evolution of SFRD and $\rhostar$ as functions of redshift. 
We decompose the total SFRD and $\rhostar$ into different galaxy mass ranges using our simulations. 
We note that the galaxy masses used for decomposition are the values at each redshift. 
At $z > 5.5$, most of the stars form in low-mass galaxies with $\Mstar = 10^7 - 10^8 \himsun$, while at $z < 3.5$ most form in massive galaxies with $ \Mstar > 10^9 \himsun$.
The same trend is also found in the $\rhostar$ evolution. 
It shows that the low-mass galaxies are preferred sites for SF in the early Universe as one would expect in the hierarchical structure formation model. 

The black solid lines in Figure~\ref{fig:sf} show the sum of SFRD and $\rhostar$ in three different galaxy mass ranges; i.e., the total SFRD and $\rhostar$ for all galaxies with $\Mstar > 10^7 \himsun$.
An important point to notice in Figure~\ref{fig:sf} is that the simulated {\em total} SFRD is well within the observed range (yellow shade and the other data points including those from gamma-ray bursts by \citet{Kistler.etal:09}), while the simulated {\em total} $\rhostar$ is clearly higher than the observed $\rhostar$. 
Since the black solid line in Fig.~\ref{fig:sf}b is a direct outcome of simulated SFRD in our simulation, the two are theoretically consistent with each other as we described at the end of Section~\ref{sec:method}. 
Whereas in the observations, different techniques are used to estimate SFRD and $\rhostar$ with different mass limits, therefore the inconsistency could arise between the two measurements. 
The comparison of black solid line in Fig.~\ref{fig:sf}b and the observed data clearly indicates the inconsistency between the observational measurements of SFRD and $\rhostar$.

According to our simulation, the solution to the above discrepancy is that the current observations are missing the starlight from low-mass galaxies at $z\gtrsim 3$, and that the low-mass end of GSMF is steeper than what current observations suggest. 
This would explain why the observed $\rhostar$ is lower than the simulation result.
This point was partly addressed earlier by \citet{Nagamine.etal:04}, but our new set of simulations reinforce their argument with a better handle on the GSMF over a wider range of galaxy mass. 

The uncertainty in the dust extinction correction needs to be addressed, as the rest-frame UV light can be extinguished by dust. 
\citet{Ouchi.etal:09} estimated the SFRD at $z\sim 7$ both with and without dust corrections as shown in the Figure~\ref{fig:sf}a, integrating their estimated UV luminosity function down to zero luminosity.  
Their total SFRD estimate (inverted solid triangle) agrees with our simulation very well. 
Therefore it is unlikely that the dust extinction correction plays a significant role in resolving the inconsistency between the observed SFRD and $\rhostar$ estimates.

Furthermore, the observational estimates by \citet{Gonzalez.etal:10} at $z>4$, which takes into account of galaxies with $> 0.3 L^{\star}_{z=3}$ show a good agreement with our simulation results for galaxies with $\Mstar \gtrsim 10^9 \himsun$ for both SFRD and $\rhostar$.
The correspondence between luminosity and stellar mass has a large scatter, but we find $L^{\star}_{z=3} \approx 10^{10}\Msun$ both observationally and in our simulations.  
Therefore the $>0.3 L^{\star}_{z=3}$ limit of \citet{Gonzalez.etal:10}  would correspond to roughly $\Mstar \sim 10^{9.5} \himsun$. 
Observational estimates by \citet{Bouwens.etal:10b} include galaxies with an order of magnitude less luminous galaxies ($> 0.05 L^{\star}_{z=3}$, corresponding to roughly $\Mstar > 5\times 10^8 \Msun$), so it would correspond to the sum of a part of red line ($10^8<\Mstar/\Msun<10^9$) and the blue line ($\Mstar > 10^9 \Msun$). 
Therefore a proper comparison between simulations and observations with the same galaxy mass limit does not show the inconsistency of our concern, and the nice agreement between simulations and the data of \citet{Ouchi.etal:09} and \citet{Gonzalez.etal:10} support our argument that the missed low-mass galaxies can account for the discrepancy between the observed SFRD and $\rhostar$. 

\citet{Jaacks.etal:2011} also demonstrated that the UV luminosity functions from our simulations and observations show an acceptable agreement, while the GSMFs show a larger discrepancy.
Generally, SFRD is estimated from the rest-frame UV measurement, and the nice agreement between the SFRDs of simulation and observations supports the agreement in the UV luminosity function as well. 
Whereas the $\rhostar$ is computed by integrating the GSMFs, which constrained less robustly at high-redshift due to uncertain light-to-mass conversion. 
This also supports our argument that the origin of the discrepancy being in the low-mass end slope of the GSMF. 

The reasonable agreement between our total SFRD (black solid line in Fig.~\ref{fig:sf}a) and the data points of \citet{Kistler.etal:09} corroborates our argument that the low-mass galaxies are the dominant contributor to the total SFRD at high redshift.  
This is because the GRB observations are independent of the rest-frame UV flux limit of galaxy surveys, and GRBs would occur wherever the star formation takes place.  
However the GRBs may also be biased tracers of star formation, and the current GRB data suggests that an enhancement of the GRB rate at high-$z$ is necessary to account for the observed redshift distribution of GRBs \citep{Virgili.etal:11}. 
A GRB rate increase due to lower metallicity at high-$z$ may not be the singular cause, and other effect such as the evolution of the GRB luminosity function break with redshift might be important.  
Given these complications, further studies are needed in the future to better understand the relationship between the GRB rate and cosmic SFRD.

\section{Discussion and Conclusion}
\label{sec:summary}

Based on the results of self-consistent cosmological hydrodynamic simulations, we argued that the current high-$z$ observations are missing low-mass galaxies with $\Mstar \lesssim 10^9\Msun$ at $z>4$ when accounting for the total $\rhostar$, and this results in the inconsistency between the observational estimates of SFRD and $\rhostar$. 
But are there any other possibilities to explain the above inconsistency, or is it possible that our simulations are incorrect?
Below we discuss two points for such possibilities. 

First possibility is that our simulation might be overproducing the low-mass galaxies at $z>4$. 
Recent observational and theoretical studies show that the low-mass, high-$z$ galaxies could have lower SF efficiencies than the normal local galaxies \citep{Wolfe.Chen:06,Gnedin.Kravtsov:10} owing to lower molecular hydrogen fractions in low-metallicity environments.  
If this is true, our current simulations might be overpredicting the $\rhostar$ at $z>4$.  
However if we revise our SF model to account for this effect, the SFRD will also decrease together with $\rhostar$, and we might underpredict SFRD instead. 
In the future we will revise our SF model to consider the H$_2$ fraction in high-$z$ galaxies, and evaluate how strong this effect would be.  

The second possibility is that the IMF at high-$z$ could be different from the local one. 
All data in Figure~\ref{fig:sf} assume the \citet{Salpeter:55} IMF.
As we emphasized in the last paragraph of Section~\ref{sec:method}, the amount of gas converted into stars is a direct output of our simulations, therefore there are no uncertainties as to stellar IMF except the instantaneous gas-recycling fraction, i.e., the ratio between the gas expelled from stars (via stellar winds and supernovae) and the total initial stellar mass. 
In our simulations, we assume the Salpeter IMF whose instantaneous gas-recycling fraction ($\beta$) is $\sim$0.1 \citep{Springel.Hernquist:03}.
If we use the \citet{Chabrier:03} IMF that has a lower number of low-mass stars, $\beta$ increases to $\sim$0.2, which would reduce the simulated $\rhostar$ by about 10\%.
Note that $\beta$ here only considers the gas-recycling from massive stars.
If we take into account of the contribution from long-lived, low-mass stars, the value of $\beta$ will increase by a factor of a few.  
However, this paper focuses on the high-$z$ galaxies, and we can safely ignore the gas-recycling from low-mass stars.  
Therefore the change from Salpeter to Chabrier IMF cannot fully account for the inconsistency between SFRD and $\rhostar$. 

The top-heavy IMF in high-$z$ galaxies has been speculated by many authors \citep[e.g.,][]{Larson:05,Fardal.etal:07,Dave:08,vanDokkum:08,Wilkins.etal:08,Bailin.etal:10}, and it has a much higher value of $\beta$. 
Consequently, the resulting $\rhostar$ will decrease significantly for a given SFRD, and it may alleviate the discrepancy in $\rhostar$.
However, the current theoretical models and observational support for top-heavy IMF are not very robust; for example, \citet{vanDokkum.Conroy:10} suggested abundant low-mass stars in high-$z$ star-forming galaxies, which argues against a top-heavy IMF. 

In summary, we find a good agreement between our self-consistent cosmological simulations and the observational estimates of SFRD and $\rhostar$ if we limit the comparison to galaxies with $\Mstar > 10^9\himsun$.  
In particular, the consistency between our simulations and the observational estimates by \citet{Kistler.etal:09}, \citet{Ouchi.etal:09} and \citet{Gonzalez.etal:10} at $z>4$ is very encouraging. 
Our simulations predict the existence of numerous low-mass galaxies with $\Mstar < 10^9\himsun$, and these low-mass galaxies are not included in the current observational estimates of $\rhostar$. 

This paper demonstrates that the current observational estimates and understanding of the formation of high-$z$ galaxies are still uncertain. 
To resolve this problem, we need future observations with increasing sensitivity and reduced uncertainties (e.g., JWST) to provide more robust constraints for the abundance of low-mass galaxies at $z>4$, as well as the improvement in the theoretical modeling of  galaxy formation, particularly on star formation and its feedback.

%% ==================================================================

\section*{Acknowledgements}
This work is supported in part by the NSF grant AST-0807491, NASA grant HST-AR-12143-01-A, National Aeronautics and Space Administration under Grant/Cooperative Agreement No. NNX08AE57A issued by the Nevada NASA EPSCoR program, and the President's Infrastructure Award from UNLV.
JHC acknowledges the support from University of Kentucky. 
This research is also supported by the NSF through the TeraGrid resources provided by the Texas Advanced Computing Center.
Some numerical simulations and analysis have also been performed on the UNLV Cosmology Cluster.

%\bibliographystyle{mn2e}
%\bibliographystyle{apj}
%\bibliography{MyRef}

\end{document}